\documentstyle[aps,multicol,epsfig]{revtex}

\newcommand{\simge}
{\raisebox{-0.75ex}[-1.5ex]{$\;\stackrel{>}{\sim}\;$}}

\def\s{{\sigma}}
\def\e{{\epsilon}}
\def\k{{ {\bf k} }}

\def\q{{ {\bf q} }}

\def\w{{\omega}}
\def\a{{\alpha}}

\begin{document}
\draft

\def\runtitle{
Origin of the Superconductivity in 
$\beta'$-(BEDT-TTF)$_2$ICl$_2$ under High Pressure
}
\def\runauthor
 {Hiroshi {\sc Kontani}}

\title{
Origin of the Superconductivity in 
$\beta'$-(BEDT-TTF)$_2$ICl$_2$ under High Pressure \\
and $\beta$-(BEDT-TTF)$_2$X at Ambient Pressure
}

\author{
Hiroshi {\sc Kontani}
}

\address{
Department of Physics, Saitama University,
255 Shimo-Okubo, Saitama-city, 338-8570, Japan.
}

\date{\today}

\maketitle      

\begin{abstract}
We present a theoretical study on the superconductivity 
in $\beta'$-(BEDT-TTF)$_2$ICl$_2$ at $T_{\rm c}=$14.2K
under high hydrostatic pressure found by Taniguchi et al., 
which is the highest record among organic superconductors.
Its electronic structure is well expressed by
the anisotropic triangular lattice Hubbard model at half filling.
In the present work,
we study this effective model
by using the fluctuation-exchange (FLEX) approximation.
In the obtained phase diagram,
the superconductivity
with $d_{x^2\mbox{-}y^2}$ like symmetry
is realized next to the antiferromagnetic (AF) insulating phase,
as a result of the 1D-2D dimensional crossover driven by the pressure.
The obtained maximum $T_{\rm c}$ is 16$\sim$18K.
In addition, the superconductivity in
$\beta$-(BEDT-TTF)$_2$X is also 
understood in the same framework.

\end{abstract}


\begin{multicols}{2}

Up to now, many kinds of organic superconductors
composed of BEDT-TTF (abbreviated as ET) molecules 
have been discovered and studied intensively.
Especially, $\kappa$-(ET)$_2$X  compounds 
attract much attention because it shows the maximum 
superconducting (SC) transition temperature ($T_{\rm c}\approx12$K)
among them
 \cite{Kanoda}.  
Recently,
the temperature-pressure phase diagram of $\kappa$-(ET)$_2$X
salt could be produced by the fluctuation-exchange (FLEX) approximation
 \cite{Kino,Kondo,Schmalian}.
It is a kind of the self-consistent spin-fluctuation theory
 \cite{Bickers,SCR}.
According to the FLEX approximation,
$d$-wave superconductivity is expected 
which is mediated by the strong antiferromagnetic (AF)
fluctuations due to the Coulomb interaction.

Quite recently,
Taniguchi et al. found the superconductivity at $T_{\rm c}=14.2$K in 
$\beta'$-(ET)$_2$ICl$_2$
under high hydrostatic pressure (P$\simge$8.2GPa),
which established the new record of $T_{\rm c}$
among the organic superconductors
 \cite{Taniguchi}.
At ambient pressure,
$\beta'$-(ET)$_2$ICl$_2$
shows a semiconducting conductivity below the room temperature,
and indicates a magnetic transition at $T_{\rm N}$=22K.
As the pressure is applied, the resistivity decreases
gradually, and the metallic behavior ($d\rho/dT >0$)
is observed above $T_{\rm MIT}$ under 6.5GPa.
Note that $T_{\rm MIT}$, which is a crossover temperature,
is expected to be higher than $T_{\rm N}$.
At 8.2GPa, insulating phase (or $T_{\rm MIT}$)
disappears and the SC transition occurs
at $T_{\rm c}=14.2$K at the same time.

According to the band calculation,
the Fermi surface (FS) of 
$\beta'$-(ET)$_2$X (X=ICl$_2$, BrICl, AuCl$_2$) 
compound is quasi one-dimensional (Q1D)
 \cite{Kato,Mori}.
It is contrastive that the FS of
$\beta$-(ET)$_2$X is almost round,
which shows the superconducting transition
at $T_{\rm c}=2.7$K for X=IBr$_2$,
at $T_{\rm c}=3.8$K for X=AuI$_2$ and
at $T_{\rm c}=1.5$K (or 8K) for X=I$_3$.
The difference of the crystal structure
between $\beta$ and $\beta'$ compounds arises from the fact
that the anion size in $\beta'$ compounds is too small to
retain the $\beta$-type structure
 \cite{Kato};
in fact, the bond lengths for I-Cl and Au-Cl
are 2.52\r{A} and 2.27\r{A} respectively,
which are smaller than the I-Br bond length
in $\beta$-(ET)$_2$IBr$_2$, 2.58\r{A}
 \cite{Kato,Mori}.

Until now,
the deformation of the structure of $\beta'$-(ET)$_2$ICl$_2$
at $P=8.2$GPa has not been determined experimentally.
In the present study, however,
we naturally assume that
its structure (gradually) approaches to that of 
$\beta$-(ET)$_2$X as the cell volume decreases
under hydrostatic high pressure, because the anion size
becomes larger compared with the cell length.
Note that the shape of the unit cells 
for $\beta$- and that for $\beta'$-structures
are almost same.
\begin{figure}
\begin{center}
\epsfig{file=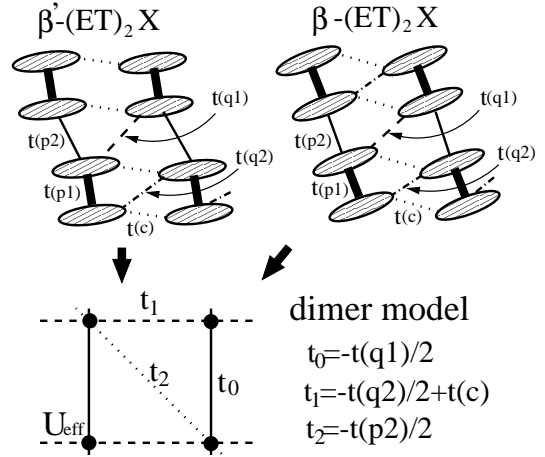,width=7cm}
\end{center}
\caption{Microscopic structures for $\beta'$-(ET)$_2$ICl$_2$
and for $\beta$-(ET)$_2$I$_3$.
They are well approximated as the 
dimer Hubbard model
($t_0$, $t_1$, $t_2$, $U_{\rm eff}$) at half filling.
In the present simplified figure,
$\beta'$-type structure is given by rotating each dimer
in $\beta$-type structure slightly in a clockwise direction.
}
  \label{fig:model}
\end{figure}
In the present work,
we study the origin of the superconductivity in
$\beta'$-(ET)$_2$ICl$_2$ on the assumption that
the crystal structure approaches to $\beta$-type
under hydrostatic pressure.
We use the FLEX approximation by noticing 
the fact that the SC phase appears by destroying
the AF phase by pressure.
Our theory can reproduce a reasonable value of 
$T_{\rm c}$ in $\beta'$-(ET)$_2$ICl$_2$ under high pressure,
as well as a low $T_{\rm c}$ in $\beta$-(ET)$_2$X
at ambient pressure.

The schematic structures of $\beta'$-(ET)$_2$X
and $\beta$-(ET)$_2$X are shown in Fig. \ref{fig:model}.
Each ellipse represents the ET molecule, and
each highest occupied molecular orbital (HOMO) of the ET molecule 
possesses 1.5 electrons on average.
The hopping parameters obtained by the band calculation
are given in Table I:
The hopping parameters for $\beta$'-(ET)$_2$AuCl$_2$
and those for $\beta$'-(ET)$_2$ICl$_2$ are similar.
Whereas, they are very different 
from the parameters for $\beta$-(ET)$_2$AuCl$_2$
although their crystal structures differ slightly.
This comes from the fact that
the overlap integral between two ET molecules
is very sensitive to the angle between them ($\phi$):
It takes the local maximum (positive) values for $\phi=0$\r{},60\r{}
and the local minimum (negative) values for $\phi=30$\r{},90\r{}
 \cite{Mori-ET}.

As a good approximation,
we take account of only the anti-bonding orbit of
each pair of the ET molecules connected by $t(p1)$,
which is more than 2.5 times larger than other hopping parameters
 \cite{HF-approx}.
Then, the original systems is mapped onto 
the ``dimer model'' at half filling,
which is the anisotropic triangular lattice Hubbard model
with four parameters ($t_0$, $t_1$, $t_2$,$U_{\rm eff}$) 
as shown in Fig. \ref{fig:model}.
Note that the $x$-$y$ coordinate in the present model 
does not correspond to that for the original systems. 
The corresponding hopping parameters
for $\beta'$-(ET)$_2$ICl$_2$ and
for $\beta$-(ET)$_2$I$_3$
are given in Table II.
\begin{center}
\begin{tabular}{|c|c|c|c|} \hline 
 & $\beta$'-(ET)$_2$ICl$_2$ \ \ 
 & \ $\beta$'-(ET)$_2$AuCl$_2$ \ 
 & \ \ $\beta$-(ET)$_2$I$_3$ \ \ 
 \\ \hline \hline
 $\ t(p1) \ $ & $-0.272$ (eV) & $-0.264$ (eV) & $-0.245$ (eV) \\  \hline
 $t(p2)$ &   $0.016$ & $ 0.020$ & $-0.084$ \\ \hline 
 $t(q1)$ &  $-0.100$ & $-0.100$ & $-0.127$ \\ \hline
 $t(q2)$ &  $-0.066$ & $-0.065$ & $-0.068$ \\ \hline
 $t(c)$  &  $-0.016$ & $-0.023$ & $ 0.050$ \\ \hline
\end{tabular}
\vspace{2mm}

{\small Table I: Hopping integrals for each systems.
They are given by the well known empirical relation $t=-10S$ (eV),
where $S$ is the overlap integral given by the band calculation
\cite{Kato,Mori}.}
\vspace{2mm}
\end{center}
\begin{center}
\begin{tabular}{|c|c|c|} \hline 
 & \ \ $x\!=\!0$ \ [$\beta$'-(ET)$_2$ICl$_2$] \ \ 
 & \ \ $x\!=\!1$ \ [$\beta$-(ET)$_2$I$_3$] \ \ \\ \hline \hline
 \ $t_0$ \  & $0.050$ (eV) & $0.064$ (eV) \\  \hline
 $t_1$ & $0.017$ &   $0.084$  \\ \hline
 $t_2$ & $-0.008$ &  $0.042$  \\ \hline
\end{tabular}
\vspace{2mm}

{\small Table II: Hopping integrals in the dimer model
for $x=0$ and $x=1$, respectively.}
\vspace{2mm}
\end{center}
Although the original on-site Coulomb interaction on a ET molecule
is $\sim 1$eV, the effective Coulomb interaction on a dimer,
$U_{\rm eff}$, is limited to $\sim 2|t(p1)| \sim 0.5$eV
 \cite{HF-approx}.

\begin{figure}
\begin{center}
\epsfig{file=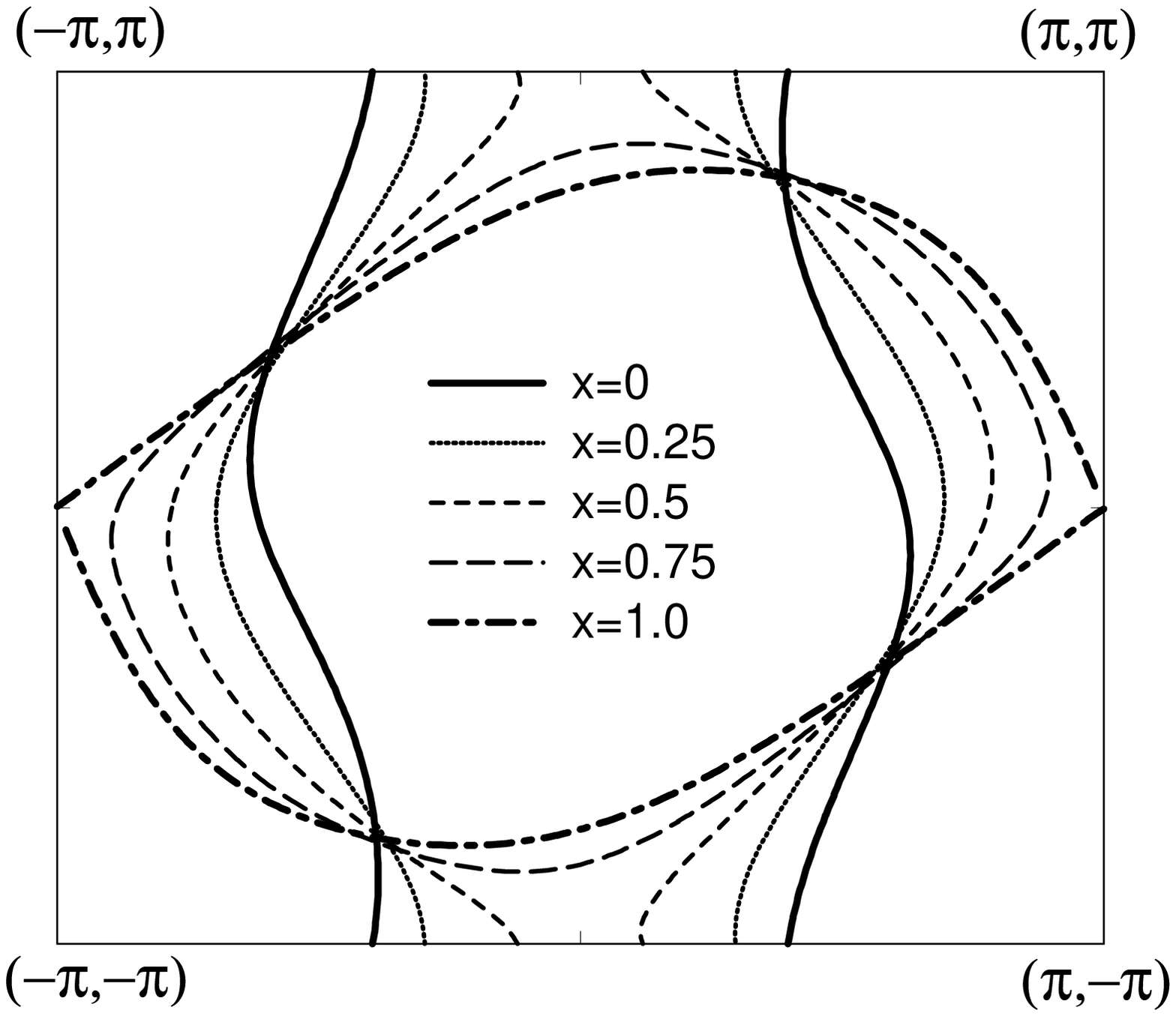,width=7.5cm}
\epsfig{file=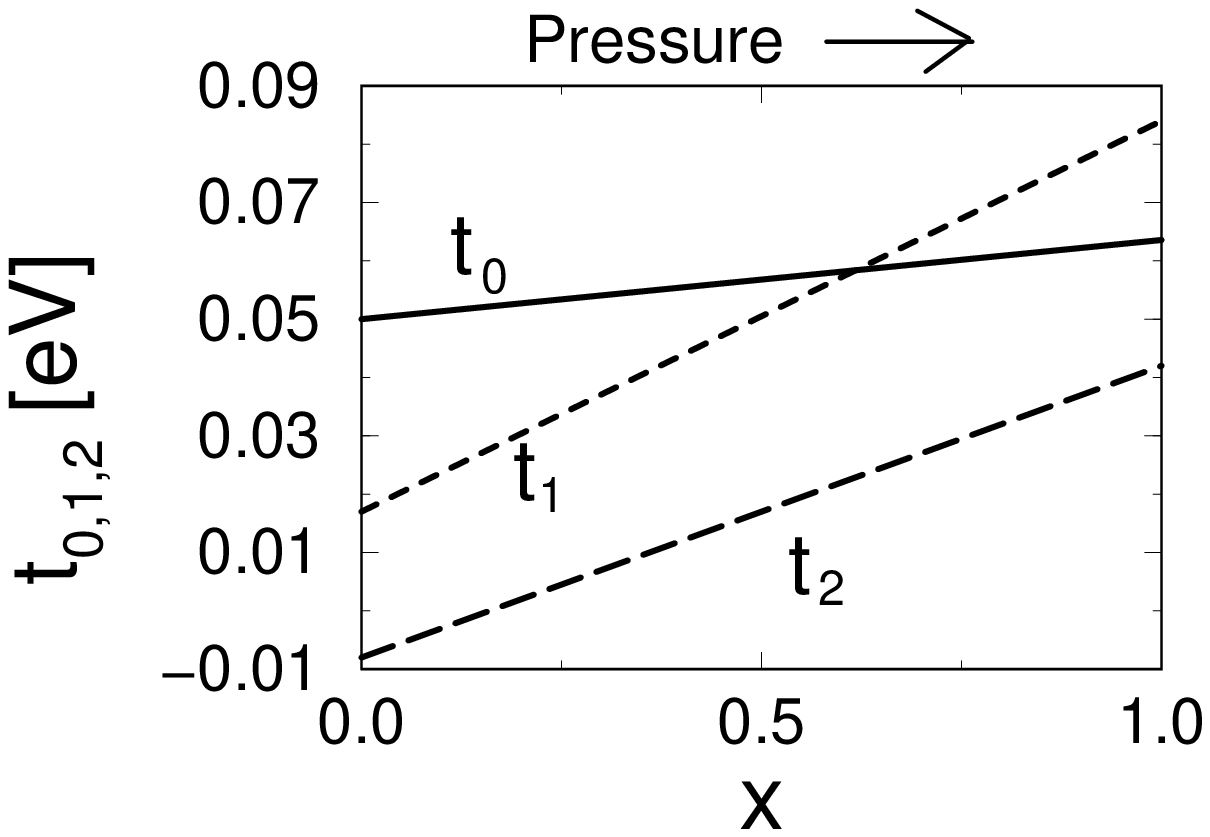,width=4.3cm}
\end{center}
\caption{
$x$ dependence of $t_0$, $t_0$, $t_0$, as well as
the corresponding FS's.
$x$ is expected to approach from $x=0$ [$\beta'$-type]
to $x=1$ [$\beta$-type]
as the applied pressure increases.
}
  \label{fig:FS}
\end{figure}

Hereafter, we study the dimer model 
given in Fig. \ref{fig:model} at half filling.
To analyze the pressure effect on the electronic states,
we put the hopping parameters of the dimer model
($t_0$, $t_1$, $t_2$) as follows:
\begin{eqnarray}
& &t_0= \ 00.5 + 0.014 x \
 \ \ \ \ \ {\rm (eV)} ,\nonumber \\
& &t_1= \ 00.17 + 0.067 x 
 \ \ \ \ \ {\rm (eV)}, \label{eqn:hopping} \\
& &t_2= -0.008 + 0.05 x 
 \ \ \ \ \ {\rm (eV)}, \nonumber 
\end{eqnarray}
where $0\le x \le 1$ is a free parameter.
$x=0$ ($x=1$) corresponds to $\beta'$-(ET)$_2$ICl$_2$
($\beta$-(ET)$_2$I$_3$); see table II.
Non-linear terms with respect to $x$ are dropped 
in eq.(\ref{eqn:hopping}).
The $x$ dependence of the FS for $U_{\rm eff}=0$ is shown 
in Fig. \ref{fig:FS}.
We see that the FS becomes round
around $x\approx 0.7$.

Here, we calculate the self-energy for the dimer model
for $U_{\rm eff}=0.4 \sim 0.5$eV
by using the FLEX approximation
 \cite{Bickers}.
Hereafter, we write $U_{\rm eff}$ as $U$
for simplicity.
To obtain the magnetic transition temperature $T_{\rm N}$,
we calculate the Stoner factor 
without vertex corrections (VC's), $\a_{\rm S}$, given by
\begin{eqnarray}
\a_{\rm S}= \max_{\k}\left\{ \ U\cdot \chi^{0}(\k,\w\!=\!0)\  \right\},
 \label{eqn:Stoner}
\end{eqnarray}
where $\chi^0(\q,\w_l)$ is given by
\begin{eqnarray}
\chi^0(\q,\w_l)
 = -T\sum_{\k, n} G(\q+\k,\w_l+\e_n) G(\k,\e_n) \mbox{,}
     \label{eqn:chi0}
\end{eqnarray}
where $G(\q+\k,\w_l+\e_n)$ is the Green function
given by the FLEX approximation, and
$\w_l$ ($\e_n$) is the Matsubara frequency for boson (fermion).
$T_{\rm N}$ is determined
by the Stoner criterion; $\a_{\rm S}= 1$.
In the FLEX approximation, however,
$\a_{\rm S}$ does not exceed 1 at finite temperatures
in two dimensional systems, which is consistent with
the Mermin and Wagner theorem.
So we determine $T_{\rm N}$ by the condition $\a_{\rm S}= \a_{\rm N}$,
where we set $\a_{\rm N}$ as $(1-\a_{\rm N})^{-1} \sim O(100)$.
The AF state will occur through the weak coupling between layers, 
$J_\perp$.
 \cite{AF-condition}

Next,
We solve the linearized Eliashberg equation 
to obtain the SC transition temperature $T_{\rm c}$.
For a singlet-pairing case
[$\phi(-\k,\e_n)= + \phi(\k,\e_n)$],
it is given by
\cite{2D-SC-Monthoux}
\begin{eqnarray}
\lambda \cdot \phi(\k,\e_n) &=& -T\sum_{\q, m}
 V(\k-\q,\e_n-\e_m)  \nonumber \\
& & \times G(\q,\e_m) G(-\q,-\e_m)
 \cdot \phi(\q,\e_m), \label{eqn:lambda}
\end{eqnarray}
where 
${V}(\k,\w_l)= 
 \frac32 U^2 \frac{{\chi}^0(\k,\w_l)}{1-U{\chi}^{0}(\k,\w_l)}
- \frac12 U^2 \frac{{\chi}^0(\k,\w_l)}{1+U{\chi}^{0}(\k,\w_l)} 
 + U$.
$T_{\rm c}$ is given by the condition $\lambda=1$.
In the FLEX approximation,
a finite $T_{\rm c}$ is obtained 
even in two dimensional systems
irrespective of the Hohenberg theorem.
However, this approximation gives reasonable $T_{\rm c}$'s for
$\kappa$-(ET) organic compounds
and high-$T_{\rm c}$ cuprates
 \cite{Kino,Kondo,Schmalian,2D-SC-Monthoux}.
Note that we could not find parameters
where the triplet pairing is dominant in the present model.

\begin{figure}
\begin{center}
\epsfig{file=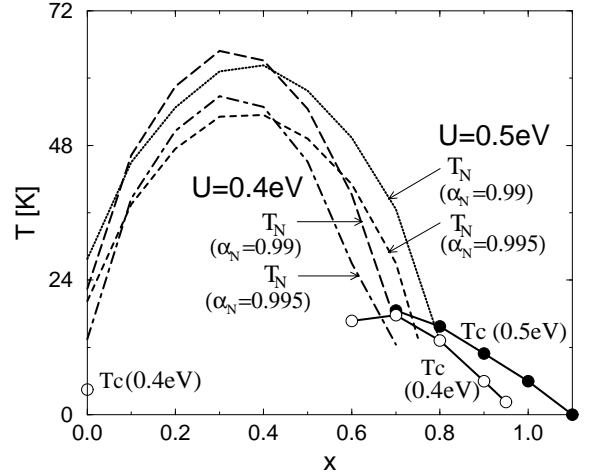,width=7.5cm}
\end{center}
\caption{
Obtained phase diagram by the FLEX approximation.
}
  \label{fig:Phase}
\end{figure}

The obtained phase diagram is given in Fig. \ref{fig:Phase}.
64$\times$64 $k$-points and 512 Matsubara frequencies are used.
The maximum $T_{\rm c}$ 
under the condition of $T_{\rm c}>T_{\rm N}$
is about 16K (18K) at $x=0.8$ ($x=0.7$)
for $U=0.5$eV ($U=0.4$eV).
$T_{\rm c}$ decreases as $x$ increases, and $T_{\rm c}\sim 6$K
at $x=1$ for $U=0.5$eV, which is consistent with the
observed $T_{\rm c}$ in several $\beta$-(ET)$_2$X compounds.
The present study also explains the low $T_{\rm c}$ in 
$\beta$-(ET)$_2$X at ambient pressure, which corresponds to
$x=1$ in Fig. \ref{fig:Phase}.
On the other hand, 
we could not find the SC phase in the metallic region for $x<0$.
We comment that $T_{\rm c}\sim4$K is obtained at $x=0$
for $U=0.4$eV as shown in Fig. \ref{fig:Phase}.
However, it should be covered with the AF phase
because of $T_{\rm N}\gg T_{\rm c}$.
Actually, the Stoner factor at 4K is extremely close to 1;
$\a_{\rm S}\approx0.999$.

On the other hand, 
the N\'eel temperature for $U=0.5$eV at $x=0$
is 28K (20K) under the condition of $\a_{\rm N}=0.99$
($\a_{\rm N}=0.995$), which is also consistent with
$T_{\rm N}=22$K in $\beta'$-(ET)$_2$ICl$_2$
at ambient pressure.
The obtained magnetic order is commensurate 
[${\vec Q}=(\pi,\pi)$] 
for $x\le0.7$ at $U=0.5$eV,
and it becomes incommensurate for $x\ge0.8$
at lower temperatures.
The obtained $T_{\rm N}$ increases as $x$ departs from 0,
which is interpreted as a natural consequence of the
dimensional increase.
It is highly demanded to experiment on the pressure dependence of 
$T_{\rm N}$.
We note that the crossover temperature  $T_{\rm MIT}$
determined by the condition $d\rho/dT=0$
will be higher than $T_{\rm N}$ because $T_{\rm MIT}$ reflects
the Mott transition, which is beyond the scope of the FLEX approximation.

\begin{figure}
\begin{center}
\epsfig{file=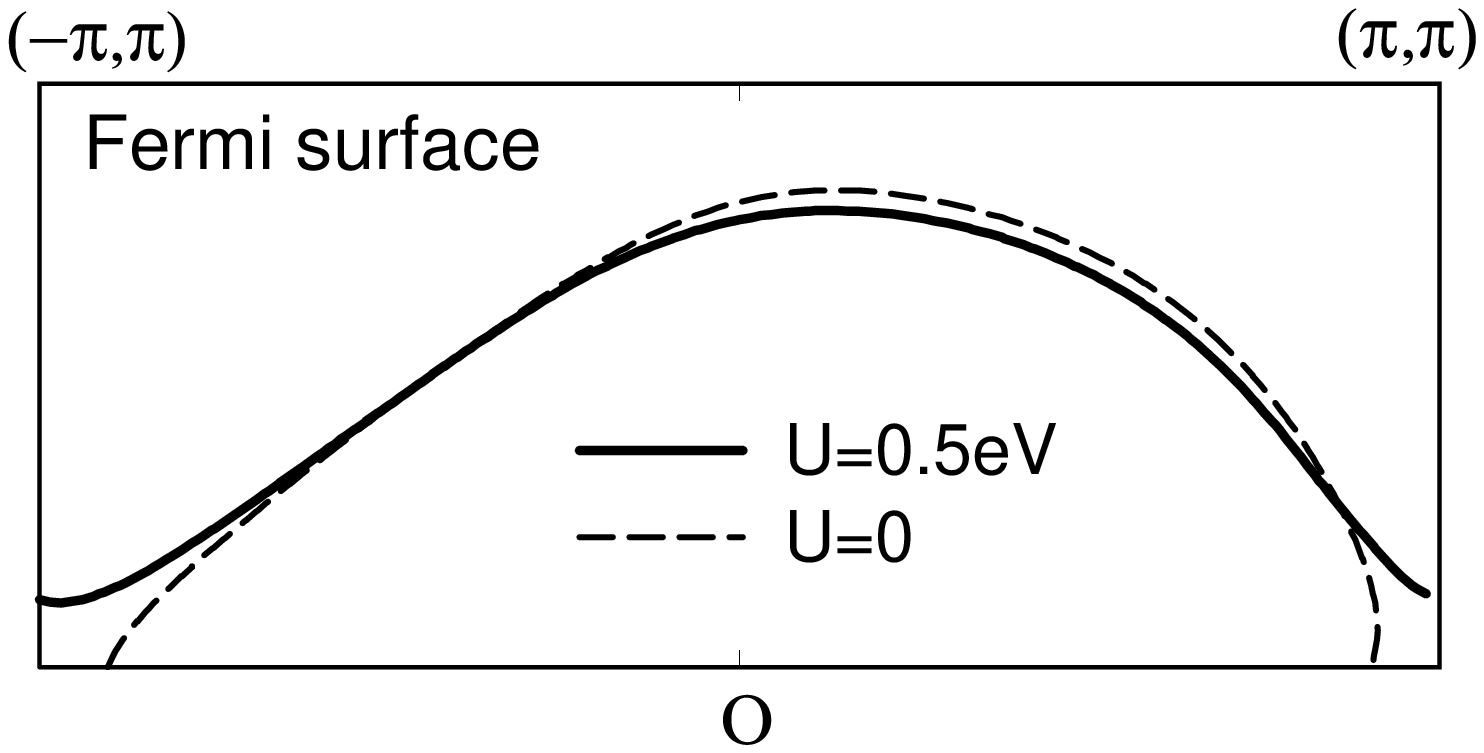,width=6cm}
\epsfig{file=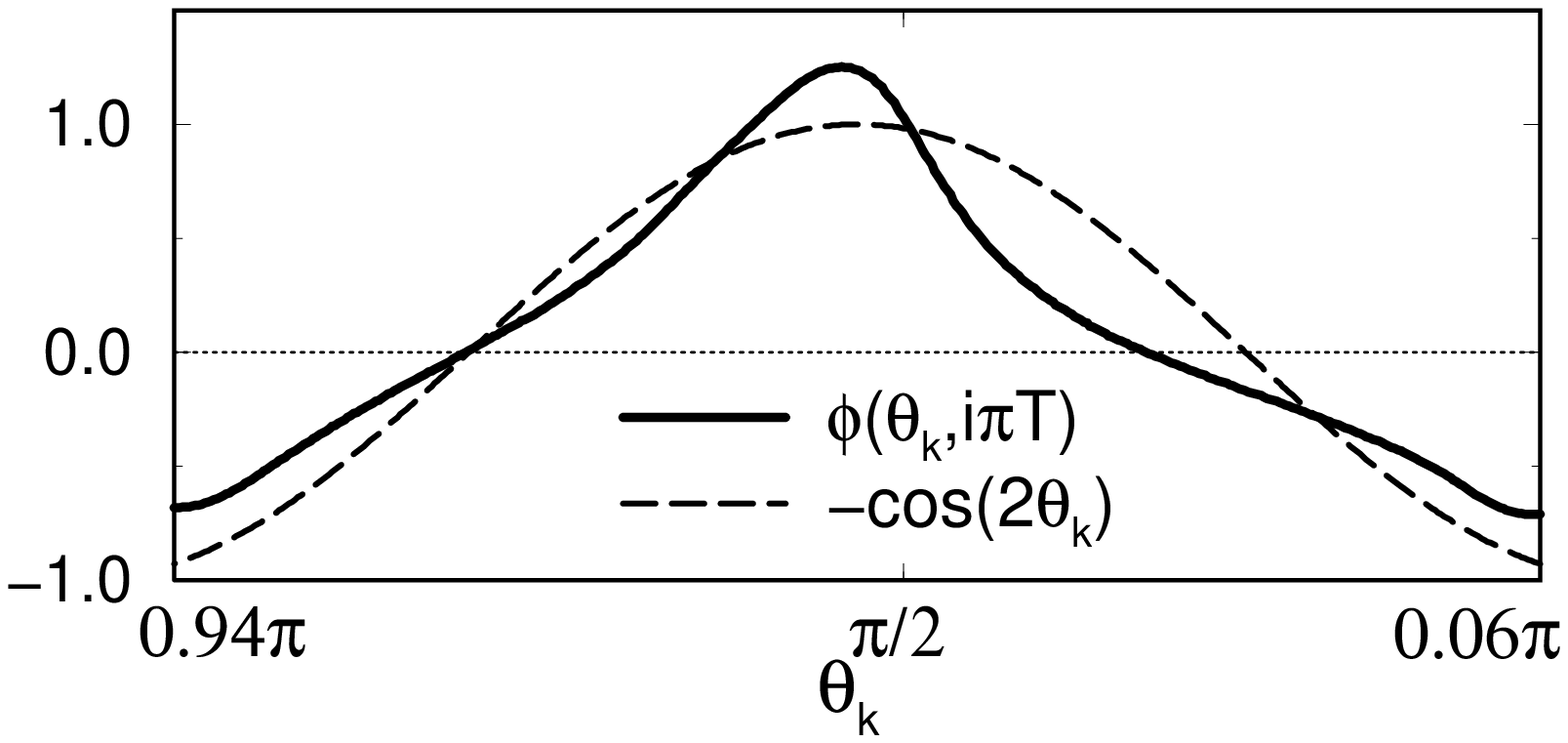,width=5.4cm}
\end{center}
\caption{
Obtained FS as well as $\phi(\theta_\k,i\pi T)$ on the FS
($U=0.5$eV, $x=0.8$, $T=15$K).
}
  \label{fig:Cph}
\end{figure}

Figure \ref{fig:Cph} shows the solution of eq.(\ref{eqn:lambda})
at $(k_x,k_y)$ on the Fermi surface, where 
$\theta_\k=\tan^{-1}(k_x/k_y)$.
Thus, the obtained SC order parameter is
$d_{x^2\mbox{-}y^2}$-wave like, as in 
the high-$T_{\rm c}$ cuprates and the $\kappa$-(ET) compounds.
Note that the present $x$-$y$ coordinate
is different from that of the original crystal;
see Fig. \ref{fig:model}.

\begin{figure}
\begin{center}
\epsfig{file=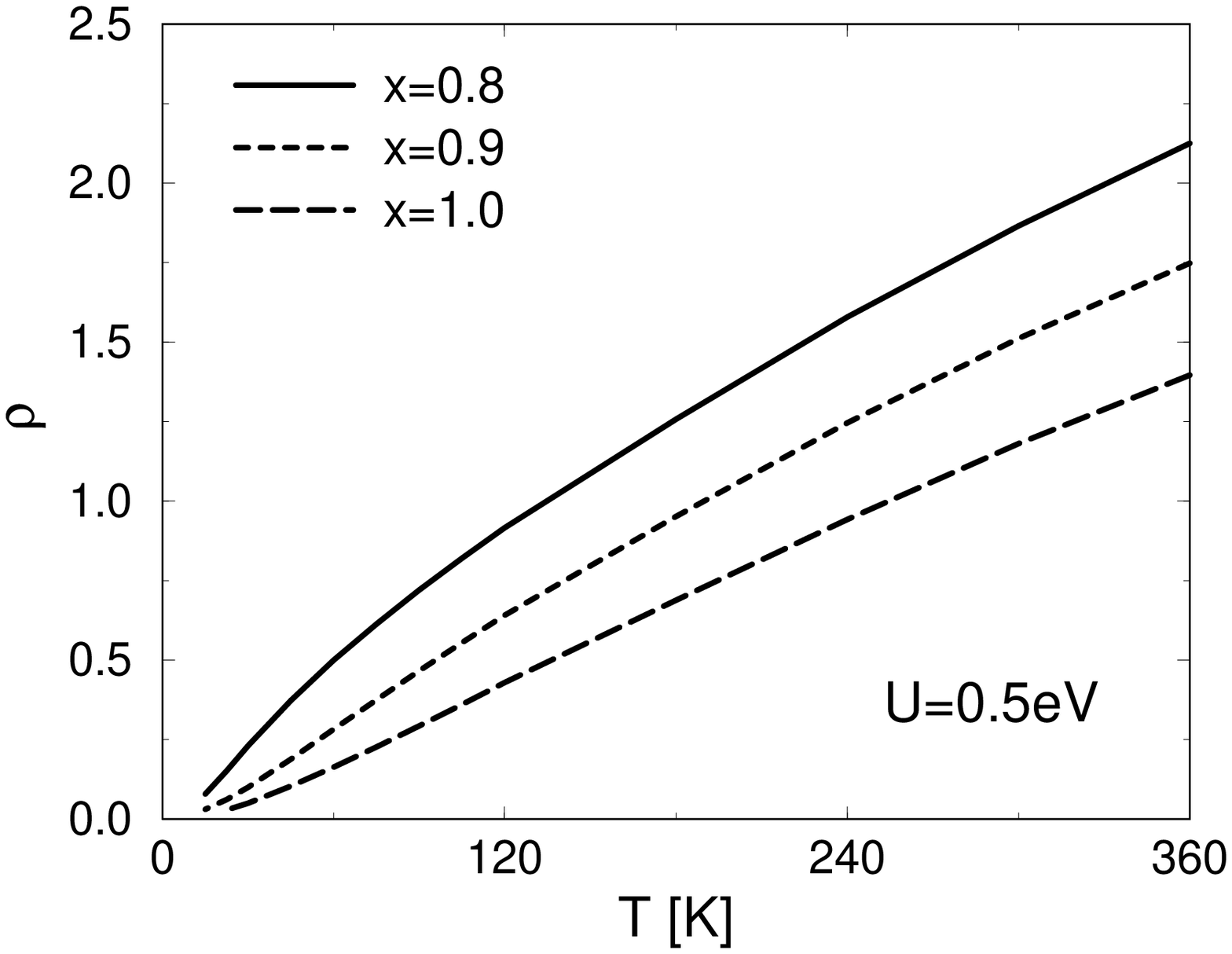,width=6cm}
\epsfig{file=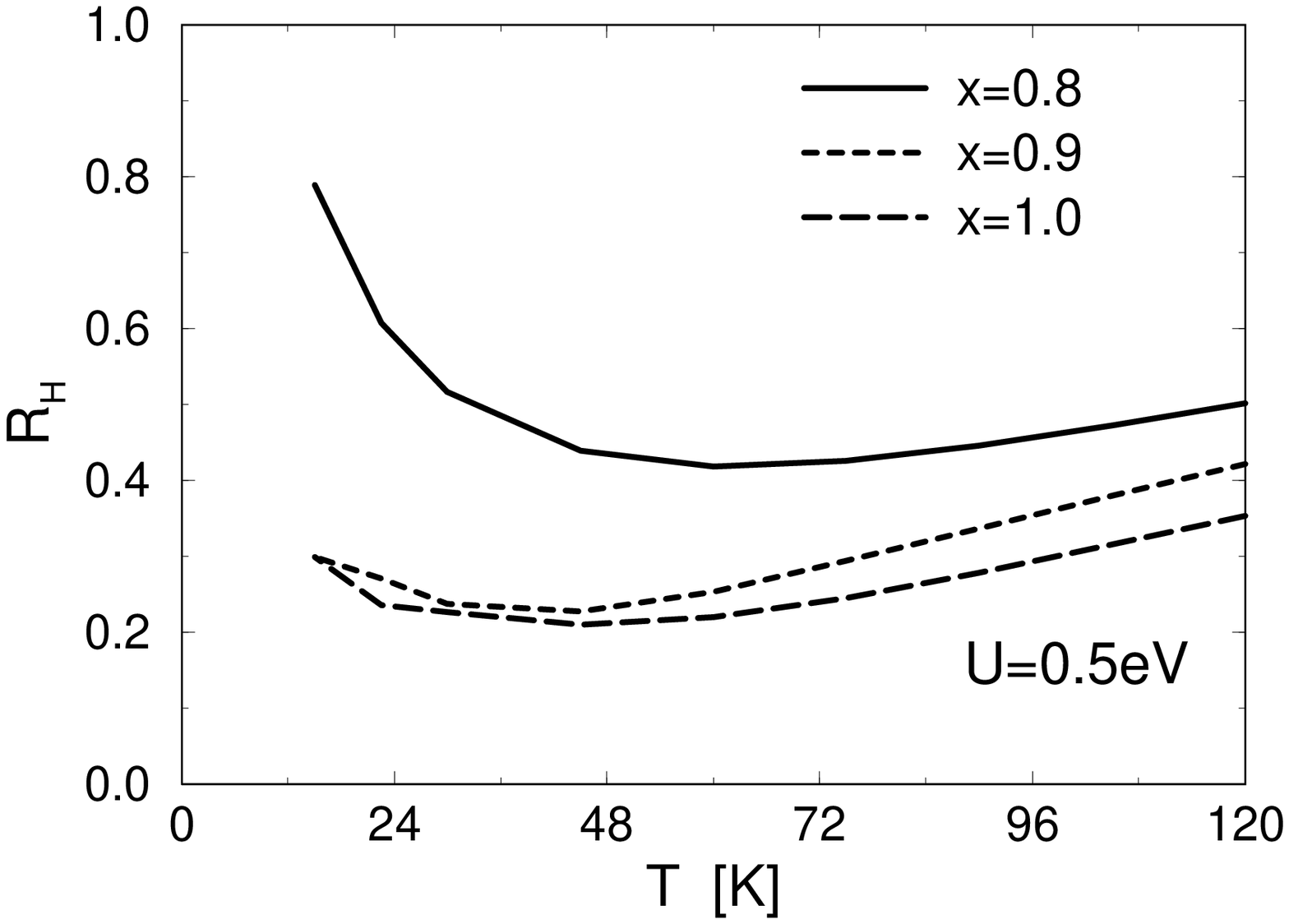,width=6cm}
\end{center}
\caption{
Obtained temperature dependences of $\rho$ and $R_{\rm H}$
with including the VC's for the current.
}
  \label{fig:Rho}
\end{figure}

Finally, we study the resistivity $\rho$ and the Hall coefficients
$R_{\rm H}$.
Considering that
the present model lacks the four-fold rotational symmetry,
we define $\rho$ and $R_{\rm H}$ as
$\rho=\sqrt{2/(\s_{xx}^2 + \s_{yy}^2)}$ and 
$R_{\rm H}= (\s_{xy}/B_z)\cdot\rho^2$ respectively:
They are independent of the choice of the $x$-$y$ coordinate.
We calculate them by including the VC's for the current
to maintain the conservation laws
 \cite{Baym}.
The role of the VC for the current,
which is totally dropped in the relaxation time approximation,
is very important in strongly correlated systems.
For example, the anomalous behaviors of $R_{\rm H}$
in high-$T_{\rm c}$ cuprates and in $\kappa$-(ET)$_2$X compounds
are reproduced only when the VC for the current
is taken into account adequately
 \cite{Hall,BEDT-Hall,MR}.
Figure \ref{fig:Rho} shows the obtained results for $U=0.5$eV.
Experimentally,
$\rho$ under 8.2GPa continues to increase up to 300K
approximately proportional to $T$,
which is similar to $\rho$ in high-$T_{\rm c}$ cuprates
 \cite{Taniguchi2}.
Such a ``bad metal'' behavior of $\rho$ is well reproduced
in the present study.
In Fig. \ref{fig:Rho},
$R_{\rm H}$ for $x=0.8$ increases
below $\sim 50$K, which is caused by the VC's for 
the current due to the AF fluctuations.
However, the obtained enhancement of $R_{\rm H}$ is 
rather smaller than that for high-$T_{\rm c}$ cuprates
or $\beta$-(ET)$_2$X compounds.
As for $x=0.9$ and $1.0$, the temperature dependence of
$R_{\rm H}$ is much moderate.
It is highly demanded to experiment on $R_{\rm H}$
in $\beta'$-(ET)$_2$ICl$_2$.
We stress that 
the observed resistivity under high pressure
is considered to be intrinsic
free from the lattice contraction effect,
which is considerably large in usual organic metals
at ambient pressure
 \cite{Taniguchi2}.

In more detail, however,
it might be better to analyze the 
original tight-binding model in Fig.\ref{fig:model}, 
instead of the dimer model, 
because a transport coefficient
is in general sensitive to the shape of the FS.
Here, we discuss the validity of the pressure induced
dimensional crossover in $\beta'$-(ET)$_2$ICl$_2$.
According to Taniguchi,
the resistivity along the less-conductive direction
(i.e., $c$-axis direction)
decreases very rapidly as the pressure increases,
which suggests the system becomes 2D-like
 \cite{Taniguchi2}.
In fact, the Q1D nature in $\beta'$-compound
is ascribed to the accidental cancellation of $t_1$
in the dimer model although $|t(q2)|$ and $|t(c)|$
are not so small; see Fig. \ref{fig:model} and Table I.
It suggests that the dimensional crossover easily occurs.
Moreover,
$\beta'$-(ET)$_2$AuCl$_2$ remains semiconducting
even at 9GPa \cite{Taniguchi2}, 
which means that the $\beta'$-type structure 
in $\beta'$-(ET)$_2$AuCl$_2$ is more robust against the pressure.
It is natural because Au-Cl bond length is about 12\% 
smaller than the I-Cl bond length.

As a result, the obtained phase diagram,
Fig.\ref{fig:Phase}, as well as the concept of
the SC driven by the dimensional crossover,
will make sense and be reasonable,
although the variable $x$ in eq.(\ref{eqn:hopping})
cannot be interpreted as the pressure simply.
The large density of states (DOS) at the Fermi level
for $x\approx0.7$ 
due to the van Hove singularity around $(\pm\pi,0)$
(see Fig.\ref{fig:FS})
accounts for the high $T_{\rm c}$
in $\beta'$-(ET)$_2$ICl$_2$.
Actually,
we could not obtain higher $T_{\rm c}$ ($\simge 5$K) 
under the condition that $T_{\rm c}>T_{\rm N}$
by the slight modification of the original parameters for 
$\beta'$-(ET)$_2$ICl$_2$,
as far as the FS is Q1D.
We note that $T_{\rm c}$ in a Q1D system, TMTSF,
is very low ($\approx1$K),
which is recognized by the FLEX approximation
 \cite{Kino-TMTSF}.

In the present analysis,
we might have underestimated
the enhancement of the ratio
$W_{\rm band}/U$ ($W_{\rm band}$ being the band width)
due to the applied pressure.
Actually,
the hydrostatic pressure not only distorts the 
structure ($\beta' \rightarrow \beta$),
but also shortens the cell volume,
which enhances each overlap integral.
Then, the enhancement of $t(p1)$,
which is proportional to $U_{\rm eff}$,
is weaker than other hopping parameters
 \cite{HF-approx}.

In summary,
we studied the origin of the superconductivity
in $\beta'$-(ET)$_2$ICl$_2$ based on the FLEX approximation,
by assuming that the structure approaches to $\beta$-type
under the hydrostatic pressure.
Our theory predicts that the 
$d$-wave superconductivity occurs 
in $\beta'$-(ET)$_2$ICl$_2$ 
as a result of the 1D-2D dimensional crossover 
owing to the crystal structure change
under high pressure.
Moreover, both $\rho$ and $R_{\rm H}$
were studied in terms of the conserving approximation.
Experimental studies on the structure of 
$\beta'$-(ET)$_2$ICl$_2$ 
as well as the band calculations 
under the condition of high pressure
are highly demanded.
In addition, the superconductivity in
$\beta$-(ET)$_2$X compound
is also understood in the same framework.

The author is grateful to
H. Taniguchi and M. Miyashita
for valuable discussions on experiments.
He is also grateful to R. Kato for useful informations
on band calculation.


\end{multicols}


\begin{thebibliography}{99}

\vspace{-15mm}

\bibitem{Kanoda} 
K. Kanoda, Physica C {\bf 282-287}  (1997) 299.

\bibitem{Kino} 
 H. Kino and H. Kontani:
 J. Phys. Soc. Jpn. {\bf 67} (1998) 3691.
\bibitem{Kondo}
 H. Kondo and T. Moriya:
 J. Phys. Soc. Jpn. {\bf 67} (1998) 3695.
%
\bibitem{Schmalian}
 J. Schmalian:
 Phys. Rev. Lett. {\bf 81} (1998) 4232.

\bibitem{Bickers} 
 N. E. Bickers, D. J. Scalapino and S. R. White:
 Phys. Rev. Lett. {\bf 62} (1989) 961.

\bibitem{SCR}
 T. Moriya and K. Ueda: Adv. Physics {\bf 49} (2000) 555.

\bibitem{Taniguchi} 
 H. Taniguchi et al.:
 J. Phys. Soc. Jpn. {\bf 72} (2003) 468.

\bibitem{Kato} 
 H. Kobayashi R. Kato and A. Kobayashi:
 Synth. Met. {\bf 19} (1987) 263.

\bibitem{Mori} 
T. Mori and H. Sasaki:
 Solid State Comm. {\bf 62} (1987) 525.

\bibitem{Mori-ET} 
T. Mori A. Kobayashi, Y. Sasaki, H. Jobayashi, G. Saito and H. Inoguchi:
 Bull. Chem. Soc. Jpn. {\bf 57} (1984) 627.

\bibitem{HF-approx} 
 H. Kino and H. Fukuyama:
 J. Phys. Soc. Jpn. {\bf 65} (1996) 2158,
 and references therein.

\bibitem{2D-SC-Monthoux} P. Monthoux and D. J. Scalapino:
 Phys. Rev. Lett. {\bf 72}   (1994)1874.

%
\bibitem{AF-condition} 
 The AF order will occur on the condition that
 $1-J_\perp \chi(Q,0)=0$. That is, $J_\perp = U(1-\a_{\rm S})\sim O$(10K),
 which will be realistic in real materials.

\bibitem{Baym}
 G. Baym and L.P. Kadanoff: 
 Phys. Rev. {\bf 124} (1961) 287.

\bibitem{Hall} 
H. Kontani et al.:
 Phys. Rev. B {\bf 59} (1999) 14723.

\bibitem{BEDT-Hall} 
H. Kontani and H. Kino:
 Phys. Rev. B {\bf 63} (2001) 134524.

\bibitem{MR}
 H. Kontani: 
 Phys. Rev. B {\bf 64} (2001) 054413.

\bibitem{Taniguchi2} 
 H. Taniguchi: private communication.

\bibitem{Kino-TMTSF} 
 H. Kino, and H. Kontani:
 J. Phys. Soc. Jpn. {\bf 68} 1481 (1999).


\end{thebibliography}
\end{document}